# Einstein and Debye temperatures, electron-phonon coupling constant and a probable mechanism for ambient-pressure room-temperature superconductivity in intercalated graphite

Evgeny F. Talantsev[1,2]

[1]M. N. Miheev Institute of Metal Physics, Ural Branch, Russian Academy of Sciences, 18, S. Kovalevskoy St., Ekaterinburg 620108, Russia
[2]NANOTECH Centre, Ural Federal University, 19 Mira St., Ekaterinburg 620002, Russia

**Abstract**

Recently, Ksenofontov *et al* (*arXiv*:2510.03256) observed ambient pressure room-temperature superconductivity in graphite intercalated with lithium-based alloys with transition temperature (according to magnetization measurements) $T_c = 330\ K$. Here, I analyzed the reported temperature dependent resistivity data $\rho(T)$ in these graphite-intercalated samples and found that $\rho(T)$ is well described by the model of two series resistors, where each resistor is described as either an Einstein conductor or a Bloch-Grüneisen conductor. Deduced Einstein and Debye temperatures are $\Theta_{E,1} \cong 250\ K$ and $\Theta_{E,2} \cong 1{,}600\ K$, and $\Theta_{D,1} \cong 300\ K$ and $\Theta_{D,2} \cong 2{,}200\ K$, respectively. Following the McMillan formalism, from the deduced $\Theta_{E,2}$ and $\Theta_{D,2}$, the electron-phonon coupling constant $\lambda_{e-ph} = 2.2 - 2.6$ was obtained. This value of $\lambda_{e-ph}$ is approximately equal to the value of $\lambda_{e-ph}$ in highly compressed superconducting hydrides. Based on this, I can propose that the observed room-temperature superconductivity in intercalated graphite is localized in nanoscale Sr-Ca-Li metallic flakes/particles, which adopt the phonon spectrum from the surrounding bulk graphite matrix, and as a result, conventional electron-phonon superconductivity arises in these nano-flakes/particles at room temperature. Experimental data reported by Ksenofontov *et al* (*arXiv*:2510.03256) on trapped magnetic flux decay in intercalated graphite samples supports the proposition.

**Keywords:** room-temperature superconductivity; ambient pressure superconductivity; Debye temperature; electron-phonon coupling constant.

# Einstein and Debye temperatures, electron-phonon coupling constant and a probable mechanism for ambient-pressure room-temperature superconductivity in intercalated graphite

## I. Introduction

Near-room-temperature superconductivity in hydrides was experimentally discovered by Drozdov *et al.*[1] in highly compressed H₃S. Over the past decade, the joint efforts of the high-pressure physics and first-principles calculations communities have led to the experimental discovery of dozens highly compressed hydride superconductors[2–45], including the ternary hydride LaSc₂H₂₄ with a zero-resistance superconducting transition temperature $T_{c,zero}(P = 262\ GPa) = 292\ K$[46]. Several new experimental techniques have been developed during this journey[47–59]. The exploration of this scientific *terra incognita* is continuing by first-principles calculations[60–66], as well as detailed characterization of the superconducting state in highly compressed hydrides[55,67–69].

While superhydrides remain the main family in which superconductivity is expected to be found at room temperature and ambient pressure[60–64,70–72], Ksenofontov et al.[73] recently reported the discovery of a new class of superconductors at ambient pressure, in which the zero-field cooling (ZFC) and field cooling (FC) magnetization curves show a transition temperature up to $T_{c,diamagnet}(P = 0.1\ MPa) = 330\ K$. These superconductors[73] are the graphite intercalated with strontium-calcium-lithium alloys.

Here, in an attempt to determine fundamental superconducting properties of this new class of superconductors, I analyzed the reported temperature dependent resistivity data $\rho(T)$ and found that $\rho(T)$ is well described by the two-component Bloch-Grüneisen equation. Deduced Debye temperatures are $\Theta_{D,1} = (240 - 340)\ K$ and $\Theta_{D,2} = (2{,}200 - 2{,}700)\ K$. Following advanced

McMillan formalism[74–76], the Debye temperature of $\Theta_{D,2} \cong 2{,}200\ K$ and transition temperature of $T_c = 330\ K$ mean that the intercalated graphite has an electron-phonon coupling constant $\lambda_{e-ph} \cong 2.15$, which is a typical value for highly compressed near-room-temperature superconducting hydrides[10,76–84].

## II. Two series resistors model

The Debye temperature $\Theta_D$ can be derived from fitting the specific heat data $C_p(T)$ to the Debye equation[85–87]:

$$C_{p,D}(T) = \gamma \times T + 9 \times R_{gc} \times N \times \left(\frac{T}{\Theta_D}\right)^3 \int_0^{\frac{\Theta_D}{T}} \frac{x^4 e^x}{(e^x-1)^2} dx \tag{1}$$

where $\gamma$ is the Sommerfeld coefficient, $R_{gc} = 8.31\ JK^{-1}mol^{-1}$ is the universal gas constant. Some research groups[88] utilized multichannel Debye equation proposed by Bouquet et al.[89]:

$$C_p(T) = \gamma \times T + 9 \times R_{gc} \times \sum_{i=1}^{M} A_i \left(\frac{T}{\Theta_{D,i}}\right)^3 \int_0^{\frac{\Theta_{D,i}}{T}} \frac{x^4 e^x}{(e^x-1)^2} dx \tag{2}$$

where $A_i$ are constants (depended from given crystalline structure and chemical composition), $M$ is a number of the channels for the Debye modes.

Wälti et al.[90] proposed a combined Debye-Einstein model, which is also in use[91,92]:

$$C_p(T) = \gamma \times T + \alpha \times C_{p,D}(T) + (1-\alpha) \times C_{p,E}(T) \tag{3}$$

where the third term represents the optical phonon-mode contributions and it is described by Einstein equation for heat capacity:

$$C_{p,E}(T) = 3 \times R_{gc} \times \left(\frac{\Theta_E}{T}\right)^2 \frac{e^{\left(\frac{\Theta_E}{T}\right)}}{\left(e^{\left(\frac{\Theta_E}{T}\right)}-1\right)^2} \tag{4}$$

For multichannel case, Eq. 3 can be rewritten as following:

$$C_p(T) = \gamma \times T + 9 \times R_{gc} \times \sum_{i=1}^{M} A_i \left(\frac{T}{\Theta_{D,i}}\right)^3 \int_0^{\frac{\Theta_{D,i}}{T}} \frac{x^4 e^x}{(e^x-1)^2} dx + 3 \times R_{gc} \times \sum_{j=1}^{P} B_j \left(\frac{\Theta_{E,j}}{T}\right)^2 \frac{e^{\left(\frac{\Theta_{E,j}}{T}\right)}}{\left(e^{\left(\frac{\Theta_{E,j}}{T}\right)}-1\right)^2} \quad (5)$$

where $A_i$ and $B_j$ are constants, $M$ and $P$ are number of the channels for the Debye modes and the Einstein modes, respectively; $\Theta_{D,i}$ is the Debye temperature of the $i$-channel, $\Theta_{E,j}$ is the Einstein temperature of the $j$-channel.

The Debye temperature $\Theta_D$ can also be derived from fitting temperature-dependent resistivity $\rho(T)$ to the Bloch-Grüneisen[93,94] equation (where the Debye temperature, $\Theta_D$, is one of free-fitting parameters):

$$\rho(T) = \rho_0 + A \times \left(\frac{T}{\Theta_D}\right)^5 \times \int_0^{\frac{\Theta_D}{T}} \frac{x^5}{(e^x-1)(1-e^{-x})} dx \quad (6)$$

where $\rho_0$, and $A$ other are free-fitting parameters. Eq. 6 is one of the approaches to determine the Debye temperature from experimental data[76–78,92,95–103]. Wiesmann et al[104] advanced this model by assuming that there is an additional conduction channel with a temperature-independent resistance $R_{sat}$ that is connected in parallel with the resistor described by Equation 1:

$$\frac{1}{\rho(T)} = \frac{1}{\rho_{sat}} + \frac{1}{\rho_0 + A \times \left(\frac{T}{\Theta_D}\right)^5 \times \int_0^{\frac{\Theta_D}{T}} \frac{x^5}{(e^x-1)(1-e^{-x})} dx} \quad (2)$$

Based on the Matthiessen rule, Varshney[105] proposed an alternative model for the $\rho(T)$ in MgB$_2$. This model is based on the assumption (which is similar to one made in Ref.[90]) that the phonon spectrum consists of two parts: an acoustic branch of the Debye type (which is characterized by the Debye temperature $\Theta_D$) and an optical mode with characteristic Einstein temperature, $\Theta_E$. In the result, the proposed equation[105] is:

$$\rho(T) = \rho_0 + A \times \left(\frac{T}{\Theta_D}\right)^5 \times \int_0^{\frac{\Theta_D}{T}} \frac{x^5}{(e^x-1)(1-e^{-x})} dx + B \times \frac{\frac{\Theta_E^2}{T}}{\left(e^{\frac{\Theta_E}{T}}-1\right)\left(1-e^{-\frac{\Theta_E}{T}}\right)} \quad (9)$$

where the third term is the Einstein resistivity model[106,107], and $\rho_0, A, \Theta_D, B, \Theta_E$ are free fitting parameters.

Because there is no additional information on the phonon spectrum in the intercalated graphite, I employed three two-series resistances models to fit the $\rho(T)$ data reported by Ksenofontov et al.[73] for graphite intercalated with lithium-based alloys:

$$\rho(T) = \rho_0 + \sum_{i=1}^{2} A_i \times \left(\frac{T}{\Theta_{D,i}}\right)^5 \times \int_0^{\frac{\Theta_{D,i}}{T}} \frac{x^5}{(e^x-1)(1-e^{-x})} dx \qquad (10)$$

this model is designated as the 2BG model, where $\rho_0, A_i, \Theta_{D,i}$ are free-fitting parameters;

$$\rho(T) = \rho_0 + \sum_{i=1}^{2} B_i \times \frac{\frac{\Theta_{E,i}^2}{T}}{\left(e^{\frac{\Theta_{E,i}}{T}}-1\right)\left(1-e^{-\frac{\Theta_{E,i}}{T}}\right)} \qquad (11)$$

this model is designated as the 2E model, where $\rho_0, B_i, \Theta_{E,i}$ are free-fitting parameters; and Eq. 9 is designated as the BGE model.

**III. Einstein and Debye temperatures in intercalated graphite samples**

**Sample I** (in the report by Ksenofontov et al.[73]) is Li-intercalated graphite sample. Figure 1 shows fits of the $\rho(T)$ data for this sample **Sample I**[73] to three models (Eqs. 9-11). Deduced characteristic temperatures and temperature dependent contributions of each term are shown in each panel of Fig. 1. The derived temperatures $\Theta_E, \Theta_{E,2}$, and $\Theta_{D,2}$ are approximately in the same range with the Debye temperature in graphene[108–111] $\Theta_D \cong (1{,}700 - 2300)\ K$. At the same time, $\Theta_D, \Theta_{E,1}$, and $\Theta_{D,1}$ are close to the reported in-plane[112] Debye temperature in the LiC$_n$ ($n$ = 6, 12) $\Theta_{D,\parallel} \cong 350\ K$.

**Sample IIId** (in the report by Ksenofontov et al.[73]) is Sr-Ca-Li-intercalated graphite sample, which showed the diamagnetic superconducting transition $T_{c,dia} = 330\ K$. Figure 2 shows fits of the $\rho(T)$ data for this sample **Sample IIId**[73] to three models (Eqs. 9-11).

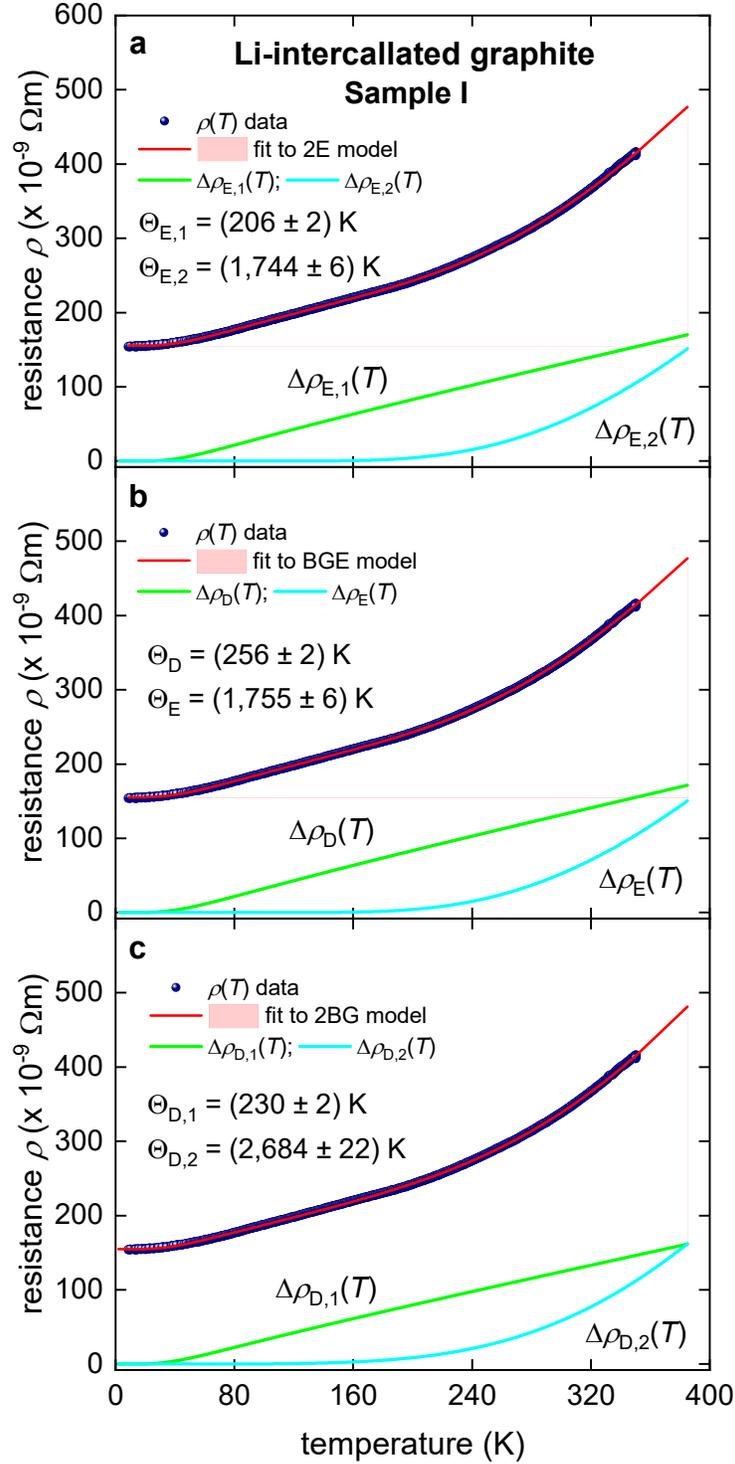

**Figure 1.** $\rho(T, B = 0)$ of the graphite intercalated with lithium (**Sample I**[73] fabricated and measured by Ksenofontov et al.[73]) and fits to (a) double Einstein model (Eq. 11, the goodness of fit $R$-square COD = 0.99992); (b) combined Bloch-Grüneisen and Einstein model[105] (Eq. 9, the goodness of fit is 0.99993); and (c) double Bloch-Grüneisen model (Eq. 10, the goodness of fit is COD = 0.99991). The thickness of 95% confidence bands (pink shadow areas) is narrower than the width of the fitting lines.

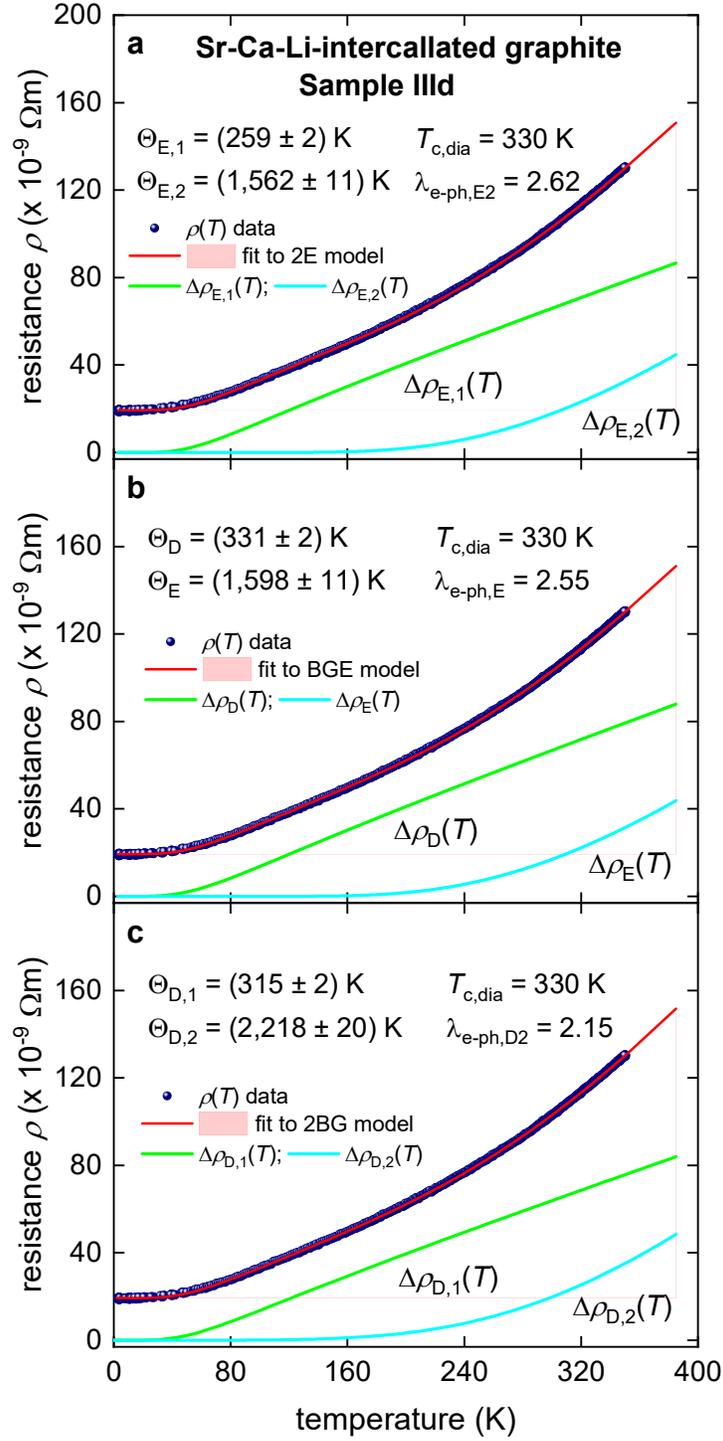

**Figure 1.** $\rho(T, B = 0)$ of the graphite intercalated with strontium-calcium-lithium alloy (**Sample IIId**[73] fabricated and measured by Ksenofontov et al.[73]) and fits to (a) double Einstein model (Eq. 11, the goodness of fit $R$-square COD = 0.99993); (b) combined Bloch-Grüneisen and Einstein model[105] (Eq. 9, the goodness of fit is 0.99994); and (c) double Bloch-Grüneisen model (Eq. 10, the goodness of fit is COD = 0.99996). The thickness of 95% confidence bands (pink shadow areas) is narrower than the width of the fitting lines.

Deduced characteristic temperatures and temperature dependent contributions of each term are shown in each panel of Fig. 2. The derived $\Theta_{E,i}$ and $\Theta_{D,i}$ values are within the same ballpark range as the values deduced in Li-intercalated graphite (**Sample I**, Fig. 1).

## IV. Electron-phonon coupling constant in room temperature superconductor Sr-Ca-Li-graphite

Deduced Einstein and Debye temperatures can be used to derive the electron-phonon coupling constant $\lambda_{e-ph}$, because the latter is the root of the advanced McMillan equations[67,74–76,113]:

$$\begin{cases} T_c = \frac{\Theta_E}{1.20} \times e^{-\left(\frac{1.04(1+\lambda_{e-ph,E})}{\lambda_{e-ph,E} - \mu^* \times (1+0.62 \times \lambda_{e-ph,E})}\right)} \times f_1 \times f_2^* \\ f_1 = \left(1 + \left(\frac{\lambda_{e-ph,E}}{2.46 \times (1+3.8 \times \mu^*)}\right)^{3/2}\right)^{1/3} \\ f_2^* = 1 + (0.0241 - 0.0735 \times \mu^*) \times \lambda_{e-ph,E}^2 \end{cases} \quad (12)$$

and

$$\begin{cases} T_c = \frac{\Theta_D}{1.45} \times e^{-\left(\frac{1.04(1+\lambda_{e-ph,D})}{\lambda_{e-ph,D} - \mu^* \times (1+0.62 \times \lambda_{e-ph,D})}\right)} \times f_1 \times f_2^* \\ f_1 = \left(1 + \left(\frac{\lambda_{e-ph,D}}{2.46 \times (1+3.8 \times \mu^*)}\right)^{3/2}\right)^{1/3} \\ f_2^* = 1 + (0.0241 - 0.0735 \times \mu^*) \times \lambda_{e-ph,D}^2 \end{cases} \quad (13)$$

where $\mu^*$ is the Coulomb pseudopotential parameter (ranging within[114] $\mu^* = 0.10 - 0.15$). In Figure 2, the derived $\lambda_{e-ph,E}$ and $\lambda_{e-ph,D}$ values were calculated in the assumption of $\mu^* = 0.10$.

## V. Discussion

Based on results reported by Ksenofontov *et al.*[73] and results reported above, I can hypothesize that the observed room-temperature superconductivity in intercalated graphite is localized in nanoscale Sr-Ca-Li metallic flakes/particles, which cannot exhibit their own phonon spectrum and thus receive the phonon spectrum from the surrounding graphite matrix (with Einstein temperature $\Theta_{E,2} \cong 1,600\ K$, or Debye temperature $\Theta_{D,2} \cong 2,500\ K$). Such high $\Theta_E$ and $\Theta_D$ bust superconducting transition temperature $T_c$ within electron-phonon phenomenology[74,75,113,115,116] (Eqs. 12,13).

In addition to these high $\Theta_E$ and $\Theta_D$, intercalated graphite has a prominent low frequency part of the phonon spectrum with $\Theta_{E,1} \cong 250\ K$ or $\Theta_{D,1} \cong 350\ K$, which boosts the electron-phonon coupling constant $\lambda_{e-ph}$:

$$\lambda_{e-ph} = 2 \times \int_0^\infty \frac{\alpha^2(\omega) \times F(\omega)}{\omega} d\omega \qquad (14)$$

where $\omega$ is the phonon frequency, $F(\omega)$ is the phonon density of states, and $\alpha^2(\omega) \times F(\omega)$ is the electron-phonon spectral function (more details can be found elsewhere[74,75,113,114,116]).

Thus, combined action of low- and high-frequency phonons causes the emergence of room-temperature superconductivity in nano-flakes/particles, because these objects cannot maintain their own phonon spectrum, and they adopt the spectrum from the surrounding graphite matrix.

Further confirmation of this hypothesis is the fact that superconducting diamagnetism is observed in ~ 0.1% of the sample volume, which can be explained by the assumption that the superconducting phase is nanoscale flakes/particles dispersed throughout the sample volume. Small flakes/particles cannot have their independent phonon spectrum from surrounding graphite matrix, and, therefore, the flakes/particles more or less adopt the phonon spectrum of the graphite. Thus, the emergence of superconductivity in these nanoscale flakes/particles depends on the match between the phonon spectrum of the graphite, density of states at the Fermi level of the flakes/particles, etc.

In overall, the superconducting state in these nano-flakes/particles can be detected by magnetic measurements (by ZFC and FC protocols), but because of the small volume fraction of these nano-flakes/particles (~ 0.1%), the percolation path[117,118] cannot be formed, and thus, the resistance measurements $R(T)$ will not show the zero-resistance state in the intercalated graphite.

Additional support for this model can be found in time-dependent magnetic measurements of the trapped magnetic flux in intercalated graphite, $m_{trap}(T)$. Truly, each nano-flake/particle exhibits room-temperature superconductivity, and, thus, integrated trapped magnetic flux of the sample should exhibit the logarithmic time dependence[119–123]. And exact this dependence was demonstrated by Ksenofontov et al.[73] (in their Fig. 4[73]). $m_{trap}(T)$ data fit to the logarithmic time dependence[119–123] showed that the decay rate is $S = 0.005 - 0.013$ [73], which is in a ballpark of typical values for other high-temperature superconductors[119–126].

## 4. Conclusions

In this paper, I analyze recently reported data[73] on $\rho(T)$ measured in an ambient-pressure room-temperature superconductor Li- and Sr-Ca-Li-intercalated graphite. The experimental data on $\rho(T)$ are well described by the model of two series resistors, and the derived values of the Einstein and Debye temperatures allow us to calculate the electron-phonon coupling constant in intercalated Sr-Ca-Li graphite: $\lambda_{e-ph,D} \cong 2.2$. The obtained value is typical for highly compressed hydride superconductors[10]. Based on the currently available data, I propose a probable mechanism for the emergence of room-temperature superconductivity in ambient-pressure intercalated graphite.

**Acknowledgements**

The author thanks Dr. V. Ksenofontov (Max Planck Institute for Chemistry, Mainz, Germany) and all co-workers of the original study[73] for providing raw experimental $\rho(T)$ data for the analysis. The work was carried out within the framework of the state assignment of the Ministry of Science and Higher Education of the Russian Federation for the IMP UB RAS. The author gratefully acknowledged the research funding from the Ministry of Science and Higher Education of the Russian Federation under Ural Federal University Program of Development within the Priority-2030 Program.